\documentstyle[12pt,epsfig]{article}

\newcommand{\be}{\begin{equation}}
\newcommand{\ee}{\end{equation}}
\newcommand{\bd}{\begin{displaymath}}
\newcommand{\ed}{\end{displaymath}}
\newcommand{\ba}{\begin{eqnarray}}
\newcommand{\ea}{\end{eqnarray}}
\newcommand{\bi}{\begin{itemize}}
\newcommand{\ei}{\end{itemize}}

\newcommand{\sbbox}[1]{\mbox{\scriptsize\bf $#1$}}

\newlength{\baselineskipsave}
\newlength{\blss}
\def\dsl{\raise.15ex\hbox{/} \kern-.57em\partial}
\def\psl{\raise.15ex\hbox{/} \kern-.57em p}
\def\ksl{\raise.15ex\hbox{/} \kern-.57em k}
\def\qsl{\raise.15ex\hbox{/} \kern-.57em q}

\begin{document}

\title{
A lattice Monte Carlo study of \\ Inverse Symmetry Breaking \\
in
a two-scalar model in three dimensions.
}

\author{G. Bimonte$^{a,b}$, D. I\~niguez $^{a}$, A. Taranc\'on  $^{a}$ and
C.L.Ullod $^{a}$ }
\bigskip
\maketitle

\begin{center}
{\it a)  Departamento de F\'{\i}sica Te\'orica, Facultad de Ciencias,\\
Universidad de Zaragoza, 50009 Zaragoza, Spain \\
\small e-mail: \tt david, tarancon, clu@sol.unizar.es} \\
{\it b)  Dipartimento di Scienze Fisiche, Universit\'a di Napoli,\\
Mostra d'Oltremare, Pad.19, I-80125, Napoli, Italy \\
\small e-mail: bimonte@napoli.infn.it\tt } \\
\end{center}
\bigskip

\begin{abstract}
{
We carry a Monte Carlo study of the coupled two-scalar $\lambda\phi^2_1
\phi^2_2$ model in three dimensions. We find no trace of Inverse Symmetry
Breaking in the region of negative $\lambda$'s for which the one-loop
effective potential predicts this
phenomenon. Moreover, for $\lambda$'s negative enough, but still in the
stability region for the potential, one of the transitions turns out to
be of first order, both
for zero and finite temperature.
}
\end{abstract}

\vskip 1cm

DFTUZ preprint 97/17~~~~~Napoli preprint 38/97~~~~~hep-lat/9707029

\newpage

\section{Introduction}

It is well known today that the strong and electroweak
interactions at the low
temperatures and energies characteristic of the present-day Universe are
well described by a spontaneously broken relativistic gauge theory.
According to this picture, a region of empty space resembles a ferromagnet
below the Curie temperature, since it is characterized by a number of
non-vanishing order parameters (one or more, depending on the elementary
particle model) which break some of its internal symmetries, an analogue of
spontaneous magnetization for ferromagnets. Another piece of wisdom, from
statistical mechanics this time, is that an ordered system, when its
temperature is raised, gradually looses its order undergoing one or more
phase transitions until it reaches a temperature above which no order at all
is present anymore.  Long ago Kirzhnits and Linde \cite{lin}
argued that
something similar might have occurred in the early Universe, when high
temperatures were present, so that the internal symmetries that appear
broken today might in fact have been  manifest at that time. Since then this
intuitive picture has been confirmed by detailed computations \cite{wei,dol}
and computer
simulations. The idea that the ground state of the Universe was symmetric in
the early times and that later, when the Universe cooled down during its
expansion, there took place a series of phase transitions that eventually
led to the present day asymmetric vacuum, is of the greatest importance for
the structure of the Universe as it influences such important issues as
baryogenesis, the formation and dynamics of topological defects \cite{kib}
like monopoles and cosmic strings, just to mention a few.

Is this scenario the only possible one? Maybe not. In fact, already in the
classic paper by S. Weinberg \cite{wei} it was pointed out
that the degree of order of
the vacuum may{\it \ increase} when the temperature is raised, contrary to
any intuition, in models with a sufficiently reach scalar sector. This
phenomenon was called Symmetry Non-Restoration (SNR) or Inverse Symmetry
Breaking (ISB), depending on whether the vacuum is ordered or disordered  at
zero temperature. It is due to the possibility that some of the scalar
fields acquire a {\it negative} Debye mass, via  {\it negative} quartic
interactions among themselves and with the other scalars, something which is
allowed in multiscalar models without causing any instability in the
potential.  To make this seemingly paradoxical statement more plausible, the
author observed that in Nature there exists at least one substance, the
Rochelle salt, which does exhibit this type of ``inverse'' behavior in a
certain range of temperatures and thus the expectation that more heat
necessarily means more disorder is clearly not always true. In any case,
these remarks of Weinberg passed totally unnoticed, until some authors
resumed this idea, applying it to important cosmological
questions like the domain wall and monopole
problems \cite{sal,goran,dvali,la}, the breaking of the CP symmetry \cite{ms},
baryogenesis \cite{dod}, inflation \cite{lee}
and the breaking of the P, Strong CP and the Peccei-Quinn
symmetries \cite{dva2}. Recently, the possibility of ISB and SNR
also in supersymmetric models has been explored \cite{rio}.

Despite the potentially important consequences for Cosmology, in the
literature on ISB and SNR there is a certain amount of scepticism about the
very existence of these phenomena. The example of the Rochelle salt is not
really persuading, because it is obvious that if  a piece of it is heated
enough it eventually melts, while in a system with ISB or SNR there is order
at arbitrarily large temperatures. Moreover the result of Weinberg was based
on a one-loop perturbative computation and so there is the possibility that
it is an artifact of this approximation. Since then, many authors have tried
to improve the one-loop result, by resorting to a number of techniques all
including some amount of non-perturbative physics. It seems fair to say that
the current situation is perplexing: while some studies have confirmed that
ISB and SNR exist \cite{bim,ame,roos,orl}, even though the region of
parameter space for which they
occur appears to be reduced in size compared to the lowest-order result,
others have come to the conclusion that these phenomena disappear once the
non-perturbative information is put in \cite{fuj,wip}. In any case, no one of
these results
appears as conclusive, because all of them rely on {\it approximations},
with various degree of applicability. The only exact result known so far
regards the lattice, where it can be shown that no order is possible at
sufficiently high temperatures for arbitrary models, gauge or not
\cite{kin}.
Unfortunately these theorems have been proven only for lattices
with finite spacing in the space-directions (but with continuum
euclidean-time axis)
and their impact for the continuum is thus unclear.

In this paper we present the results of the first Monte Carlo study of ISB.
We have considered a two-scalar model in 2+1 euclidean dimensions, for which
a one-loop computation predicts the possibility of ISB and SNR. We have
studied its phase diagram for various choices of the coupling constants as a
function of the temperature. Concretely, we have simulated the model
on various asymmetric lattices $N_t \times N_s^2$, where
$N_t$ and $N_s$ are respectively the number
of sites in the ''euclidean time'' and space directions,
the temperature being related to $N_t$ by $T=1/(a N_t)$, with $a$ the lattice
spacing. In order to avoid finite-size effects we have taken the thermodynamic
limit $N_s \rightarrow \infty$ (for any fixed $N_t$) using Finite Size
Scaling (FSS) techniques.
The high level of precision required to clearly separate the transitions
corresponding to the various values of $N_t$ required very large
statistics and correspondingly long simulation times.
We have found that
for all the values of $N_t$ that we have considered,
ISB seems ruled out, irrespective
of the values of the couplings. In fact, it appears that the size of
the disordered
region of the phase diagram {\it increases} when the temperature is raised
(i.e. when $N_t$ is reduced), as it
happens in normal cases (for example in the  $\phi^4_3$ model \cite{bitu})
, and does not decrease as required for ISB to take place.
As for SNR the results of our simulations show that, if one starts
at $T=0$ from an ordered phase, the system
disorders above a certain critical temperature $T_c$ (coherently with
the results of \cite{kin}), again as it happens in normal cases.
In principle, this does not exclude the possibility of SNR, because
it might happen that $T_c$ diverges in the continuum limit. In order
to see if this possibility really occurs, one would have to carry
 a detailed study of the continuum
limit, something that we have not done.

Another interesting result of our simulations is that one of
the transitions turned out to be of {\it first} order, both
at $T=0$ and at finite temperature, when the coupling among the
two fields is negative and large in absolute value, being of second
order otherwise. This result is in
qualitative agreement with refs.\cite{born,born2}, where
our model is considered for the particular case
when an extra $\phi_1 \leftrightarrow \phi_2$ symmetry is present
(this case of course excludes the possibility of ISB or SNR).
Using the method of the Average Effective Action \cite{wet}, the
authors found in the phase diagram surfaces of first order
phase transitions, but there this occurred
for all negative values of the coupling among the two scalars,
while we seem to find
this behavior only for couplings strongly negative.

The organization of the paper is the following. In Sec. 2 we present
the continuum model,
and discuss the main features of its lattice version, while in Sec. 3
we review the predictions of perturbation theory on its high-temperature
behavior. The simulation,
the techniques used and the results obtained are presented in Sec. 4.
Finally Sec. 5 is devoted to the conclusions.

\section{The model and its lattice formulation}

We consider the theory for two real scalar fields in 3 euclidean
dimensions, described by the bare (euclidean) action:
\be
S=\int d^3 x
\Big\{\frac{1}{2} \sum_{i=1,2}\Big[
(\partial_{\mu}\Phi_i)^2 + \frac{1}{2}m^{(i)2}_{B}
\Phi_i^2 + \frac{g_i}{4!} \Phi^4_i\Big]+
\frac{1}{4}
g\Phi_1^2\Phi_2^2 \Big\}.
\label{bact1}
\ee
What will be essential for ISB and SNR, in the above action the
quartic coupling $g$ can be {\it negative}, as the
condition of boundedness from below of the potential is satisfied if:
\ba
g &<& 0 ~,\nonumber \\
g_1 \; g_2 &>& 9\; g^2,~~~g_1> 0,~~~g_2> 0.
\label{cotas}
\ea
As it is well known scalar models with quartic couplings, like (\ref{bact1}),
in three dimensions are superrenormalizable. In the
perturbative expansion of the zero-temperature Green's functions there
is only a finite
number of primitively divergent diagrams. Depending on the regularization
scheme, UV divergencies are found only in two classes of diagrams:
the tadpole diagrams, at one loop, and the
sunset diagrams (for zero external momenta), at two loops,
both contributing to the self-energies.
As a consequence, in (\ref{bact1}) the only quantities that require an
infinite renormalization are the bare masses $m^{(i)2}_{B}$,
while the fields and
coupling constants can be identified
with the renormalized ones. It is for this reason that in
(\ref{bact1}) we have written the latter without the suffix $B$.

When regularized on an infinite three-dimensional cubic
lattice of points
$\Omega$ with lattice spacing $a$, the above action is replaced
by its discretized version
\ba
S_L&=& \sum_{x \in \Omega}   a^3 \Big\{ \sum_{i=1,2}\Big[
\sum_{\mu}\frac{1}{2}
(\Delta^{(a)}_{\mu}\Phi_{i,L})^2(x)+\frac{1}{2} m_B^{(i)2}\Phi_{i,L}^2(x)+
\frac{g_i}{4!}\Phi^4_{i,L}(x) \Big]+ \nonumber\\
&& \;\;\; \;\;\;\;\;\;\;\;\; + \frac{g}{4}\; \Phi_{1,L}^2(x)
\Phi_{2,L}^2(x)\Big\},\label{lat1}
\ea
where $\Delta^{(a)}_{\mu} $ is the lattice derivative operator in the
direction $\mu$:
\be
\Delta^{(a)}_{\mu}\Phi_{i,L}(x)\equiv \frac{\Phi_{i,L}(x +
\hat{\mu} a)-\Phi_{i,L}(x)}{a}\;.\nonumber
\ee
We find it convenient to measure all dimensionful quantities in (\ref{lat1})
in units of the lattice spacing; thus we define:
\be
\phi_{i,L}(x) \equiv a^{1/2} \Phi_{i,L}(x),~~~u_i \equiv a g_i,~~~
r_i \equiv a^2 m^{(i)2}_B~~~u \equiv a g~~.\label{BBpar}
\ee
In terms of the dimensionless quantities the lattice action now reads:
\ba
S_L&=& \sum_{x \in \Omega}  \Big\{ \sum_{i=1,2}\Big[
\sum_{\mu}\frac{1}{2}
(\Delta^{(1)}_{\mu}\phi_{i,L})^2(x)+\frac{1}{2} r_i\phi_{i,L}^2(x)+
\frac{u_i}{4!}\phi^4_{i,L}(x) \Big]+ \nonumber\\
&&\;\;\;\;  \;\;\;\;\;\;+ \frac{u}{4}\; \phi_{1,L}^2(x)
\phi_{2,L}^2(x)\Big\}\;,\label{lat2}
\ea
The standard lattice notation is obtained with a further redefinition
of the fields and couplings in (\ref{lat2}) according to:
\be
\phi_{i,L}(x) = {\sqrt \kappa_i}
\phi_{i,\sbbox{r}},~~~u_i=\frac{24 \lambda_i}{\kappa_i^2},~~~
r_i=2\frac{1-2\lambda_i -3 \kappa_i}{\kappa_i}~~
\ee
and
\be
u=\frac{4 \lambda}{\kappa_1\kappa_2}.
\ee
After these redefinitions we get our final form of the lattice action
\ba
S_{L}&&=  \sum_{\sbbox{r} \in {\bf Z^3}}\Big\{ \sum_{i=1,2} \Big[
       -\kappa_i  \sum_{\mu}\phi_{i,\sbbox{r}} \phi_{i,\sbbox{r}+
\hat{\sbbox{\mu}}}+
       \lambda_i(\phi^2_{i,\sbbox{r}}-1)^2+\phi^2_{i,\sbbox{r}} \Big]+
\nonumber\\
&&~~~~~~~~~~~~~~~~~~~+\lambda\phi^2_{1,\sbbox{r}}\phi^2_{2,\sbbox{r}}\Big\}
\label{bbfin}.
\ea

For generic values of the parameters,
the model has a ${\bf Z}_2 \times {\bf Z}_2$ symmetry which
can be spontaneously broken. We
now briefly comment on the fixed-point structure of the model.
Besides the Gaussian fixed point, corresponding to $r_1=r_2=u_1=u_2=u=0$
(or $\kappa_1=\kappa_2=1/3$ and $\lambda_1=\lambda_2=\lambda=0$), which has
a  null attractive domain in the infrared, there are five more fixed points.
They are all attractive in at least one
direction and correspond to:

\noindent
1) the Heisenberg fixed point, for $u_1=u_2= 3 u=u^*_H$ and
$r_1=r_2=r^*_H < 0$,
where the symmetry of the model is enhanced to $O(2)$.

\noindent
2) the three Ising fixed points, for $u=0$ and $u_i=u_I^*$, $r_i=r^*_I$
for some $i$,
where the model splits into two independent $\phi^4_3$ models.

\noindent
3) the Cubic fixed point, for $u_i=u=u_I^*/2$, $r_i=r^*_I$, which again
splits in two independent  $\phi^4_3$ models, after a $\pi/4$-rotation
of the fields.

Due to the rich fixed point structure, our two-scalar model exhibits
complicated
cross-over phenomena. For a study of these aspects (for the particular case
when an extra $\phi_1 \leftrightarrow \phi_2$ symmetry is present), we
refer the reader to ref.\cite{born}.

We now briefly discuss the continuum limit of the lattice model
(\ref{lat2}), even if in our simulations we have not taken this step because
the results we got seem to rule out the possibility of ISB a priori,
even in the continuum. Taking the continuum
limit would instead be necessary to prove that neither SNR occurs,
a possibility that we cannot exclude in principle, even if it
appears unlikely.

In order to go to the continuum, one has to move the
simulation point along (zero temperature)
renormalization group trajectories of the lattice model, called curves
of constant physics
(CCP). They are parametrized by the lattice spacing $a$ and are
such that any observable, measured in physical units,
approaches a definite limit when $a \rightarrow 0$. The CCP's
that one has to pick in order to
explore the models described by perturbation theory in the
continuum are those that approach the Gaussian fixed point.
This is rather evident if one recalls that in perturbation theory
the bare coupling constants $g_i$ and $g$ are finite and so, by
(\ref{BBpar}), the lattice
coupling constants $u_i$ and $u$ all vanish in the limit of
the lattice spacing going to zero.
We conclude these remarks by observing that, due
to the superrenormalizability of the model, the
asymptotic form of the CCP can be computed exactly,
by means of a simple two-loop
computation. For more details, in the case of the pure
$\phi^4_3$ model, we refer the reader to ref.\cite{bitu}.

\section{High-T perturbation theory}

In this Section we briefly review the predictions of perturbation theory
on the high-temperature behavior of the model (\ref{bact1}).
This turns out to be not an easy task.
In fact, while the breaking of perturbation
theory at high temperature is a generic feature of quantum field theories,
in three dimensions things are worse due to the presence of
severe infrared divergences. Let us see how
this comes about. The issue we are interested in is the possibility
that in our two scalar model the symmetric vacuum of the theory is
unstable at arbitrarily high temperatures, for opportune choices
of the parameters. For this
we need compute the leading high-$T$ contribution to the
second derivative of the effective potential in the origin, or
equivalently to the 1PI 2-point functions
for zero external momenta $p$.
An easy one-loop computation (in the imaginary time formalism)
\cite{kap}
in our model gives, in the limit $T/m \gg 1$, the result:
\be
M_i^2(T) \equiv -\Gamma^{(2)}_i(p=0,T)= \frac{T}{4 \pi}(g_i + g)\log
\left(\frac{T}{\mu}\right)~~,\label{selft}
\ee
where $\mu$ is the mass scale. Recalling now that the coupling $g$ can be
{\it negative}, it is easy to see that for $g_1/g_2$ sufficiently large,
the range (\ref{cotas}) of stability for the potential includes values of
$g$ such that, say, $g_2 + g$ is {\it negative} (it can be proven easily
form eqs.(\ref{cotas}) that one cannot have $g_1+g<0$ at the same time). If
we can trust eq.(\ref{selft}), it thus appears that, for $g_2+g <0$,
$M_2^2(T)$ becomes negative at sufficiently high temperatures, irrespective
of its value at $T=0$. This means that above a sufficiently high
temperature the symmetric vacuum will become unstable in the direction of
$\phi_2$ and thus there will be symmetry breaking for this field. This is
the essence of the phenomena of SNR and ISB: in multiscalar models some
Debye masses can become negative and so one can have spontaneous symmetry
breaking at arbitrarily high temperatures.

Can we trust this result in our model? As we pointed out earlier the
perturbative expansion of the effective potential in three dimensions
suffers from hard infrared divergencies that can cause its breakdown at
high temperatures. A closer look at eq.(\ref{selft}) raises the doubt that
indeed this is the case. According to eq.(\ref{selft}), we see that for
$g_2+g<0$, $\phi_2$ acquires at high temperatures a vev $\langle \phi_2
\rangle$ of order $\sqrt{T}$ (upto logarithms) . 
Now, simple power counting arguments
suggest that the loop expansion of the effective potential for $\phi_2$ is
reliable only if
\be
\frac{g_2 T}{m^{(2)2}+ \frac{1}{2}g_2\langle \phi_2
\rangle^2} \ll 1~,
\ee
in the range of temperatures and fields of interest. For
$\langle\phi_2\rangle^2\approx T$ this quantity is of order 1 and thus it
is clear that we cannot trust perturbation theory in this case.

Such a conclusion is further reinforced if we look at the effect of the
inclusion of higher order corrections. Indeed, at high $T$, powers of $T$
can compensate for powers of the coupling constants, thus producing large
corrections to the one-loop result (for an analysis of this problem in the
$\phi^4_3$ theory see the first of ref.\cite{wip}). The standard solution
to this difficulty is the resummation of the so-called ring-diagrams
\cite{dol}, which leads to a self-consistent gap equation for the Debye
masses $M_i^2(T)$. In our case the gap equations obtained in this way are:
\ba
M_1^2(T)&=& m_{1}^2+\frac{T}{4 \pi}\left(g_1 \log
\frac{T}{M_1(T)}+ g \log
\frac{T}{M_2(T)}\right)~~,\nonumber\\
M_2^2(T)&=& m_{2}^2+\frac{T}{4 \pi}\left(g_2 \log
\frac{T}{M_2(T)}+ g \log
\frac{T}{M_1(T)}\right)~~,\label{gap}
\ea
where $ m_{i}^2$ denote the renormalized masses. These equations always
admit real positive solutions for $M_i^2(T)$, for sufficiently high $T$
\cite{wip}, which means that the symmetric vacuum is stable at high
temperatures. Thus, the inclusion of higher order effects (in fact the use
of gap equation is a non-perturbative procedure, as it implies an infinite
resummation) has reversed the one-loop result.  Notice though that,
differently from what happens in four dimensions \cite{dol}, the gap
equations (\ref{gap}) do not make sense for  $M_i^2=0$ and thus, if there
is spontaneous symmetry breaking at $T=0$, they cannot be used to give an
estimate of the critical temperature $T_c$ above which symmetry is
restored. This is another consequence of the infrared singularities alluded
to before (for the case of a  $\phi^4_3$ model, the limits on the
possibility of determining $T_c$ in perturbation theory are discussed in
\cite{ein}, while a  measurement of $T_c$ as a function of the renormalized
parameters by a Monte Carlo simulation is given in \cite{bitu}). This state
of affairs is obviously unsatisfactory: perturbation theory cannot give any
information in the study of ISB and SNR in three dimensions and this
motivated us to perform a lattice Monte Carlo simulation of this model to
have a fully non-perturbative approach to this problem.

\section{The simulation}

As we saw in the previous section, according to a one loop computation
of the Debye masses, at high temperatures the symmetric vacuum of our
system should be
unstable in the direction of, say, the field $\phi_2$ for negative
values of $g$ such that  $g_2 +g < 0$.
This means that, for such values of $g$, the field
$\phi_2$ should necessarily have a non-vanishing vacuum expectation
value (vev) at high temperature.

Let us see how one can analyze this phenomenon on the lattice. To study
the effects of temperature on the system, one simulates it on asymmetric
lattices $N_t \times N_s^2$, $N_t$ and $N_s$ being the number of sites
in the euclidean time and space directions respectively. The physical
temperature is then related to $N_t$ as $T=1/(a N_t)$, $a$ being the
lattice spacing. The $T=0$ case, corresponds to symmetric lattices
$N_s^3$. In
the next step one takes the thermodynamic limit
$N_s \rightarrow \infty$, holding $N_t$ fixed:
for each $N_t$, in this limit the phase diagram of the system splits
into four
phases, distinguished by the vev's of the fields and
divided by surfaces of phase transitions. Finally, one
takes the continuum limit $a\rightarrow 0$, which implies, as
explained in section 2, moving the simulation point along some CCP
approaching the gaussian fixed point. In order to see if ISB occurs,
one now picks a CCP that describes a continuum
theory with a symmetric vacuum at $T=0$. Such a CCP must lie entirely in the
disordered region of the $T=0$ phase diagram. One has
ISB if this CCP crosses the $\phi_2$ transitions line,
separating the symmetric phase from the ordered phase with
$\langle \phi_2 \rangle \neq 0$, of some finite $N_t$ lattice.
It is obvious that this can happen only if the
$\phi_2$
transition line of some lattice with finite $N_t$ lies
in the disordered region of the $T=0$ phase diagram.
If $\bar a$ is the lattice spacing corresponding to the crossing
point, one concludes that the physical temperature
for the $\phi_2$-ISB-phase transition is $T_c=1/(\bar a N_t)$. In order
for this transition to survive in the continuum limit, our CCP
must in-fact cut all the $\phi_2$ transitions lines for sufficiently
large $N_t$'s
and it must so happen that the corresponding critical temperatures
approach a finite value for $a \rightarrow 0$.

It is clear, from the above considerations, that the first thing to do
in order to study ISB is to draw the phase diagrams of
our model, for various values of $N_t$ (including of course
for symmetric lattices).
With the parametrization
of (\ref{bbfin}) the region of ISB for $\phi_2$ corresponds to
\be
\lambda + 6 \frac{\kappa_1}{\kappa_2}\lambda_2 < 0.\label{inst}
\ee
Our model contains five independent parameters, too many
for a complete analysis to be possible. Thus, we decided
to fix the values of $\lambda_1$ and $\lambda_2$.
To choose them in an optimal way, it was necessary to make a
compromise between a number
of conditions. First, we took them such that $9\lambda_2<\lambda_1$.
This is because the stability condition eq.(\ref{cotas}) and the
one-loop condition for ISB eq.(\ref{inst}) together imply
\be
\lambda_1 > 9 \left(\frac{\kappa_1}{\kappa_2}\right)^2 \lambda_2
\ee
and near the gaussian point $\kappa_1 \approx \kappa_2 \approx 1/3$.
Another condition was that $\lambda_1$ and $\lambda_2$
had to be small in order to be near the gaussian point, but not
too small because then it would be difficult
to distinguish the various phase
transitions one from the other. Moreover,
it was desirable to have $\lambda_1$ as large as possible, in order for
the stability condition eq.(\ref{cotas})
and the ISB condition eq.(\ref{inst})
to be satisfied by a wide range of values for $\lambda$'s.
The outcome  of these considerations were the values $\lambda_1=0.06$
and $\lambda_2=0.002$. The
condition of stability for the potential then gives a lower bound
on
$\lambda$ of $\approx -0.0219$, while the one
loop calculation predicts ISB for $\lambda\leq -0.012$ (
assuming $\kappa_1 \approx \kappa_2$).
Inside the interval given
by these two bounds, we have selected two values of $\lambda$. The
first is $\lambda=-0.016$ which is well inside the stability
region but not too close to the uncertain upper bound
for the onset of ISB given by the
one loop calculation. The second is $\lambda=-0.020$: it is the
one closest to the lower stability bound that we could use, without running
into problems due to the nearby instability region.

\begin{figure}[t!]
\epsfig{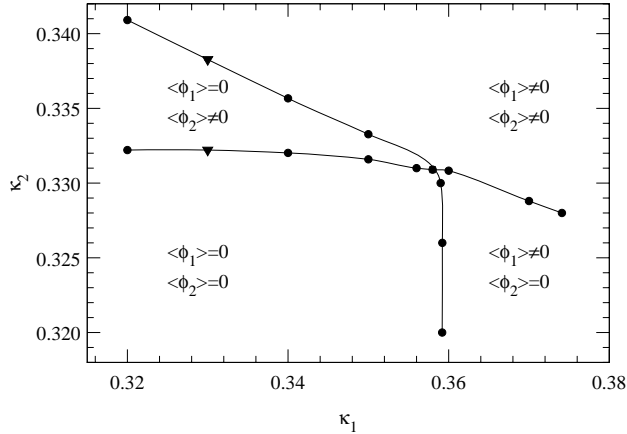}
\caption{Phase diagram at $\lambda_1=0.06,~
\lambda_2=0.002,~\lambda=-0.016$
for a $8^3$ lattice. Error are smaller than the size of the marks.
The points marked with triangles, for $\kappa_1=0.33$, are
those for which we carried the detailed study of ISB.}
\label{PH_DIA}
\end{figure}

In the simulations, we used a Metropolis algorithm
combined with a Wolff single cluster method
(the latter updates the sign of the fields), generally taking a ratio
of 20 clusters every 3 Metropolis iterations (at the points where
we found a first order transition the clustering was not efficient).
The job was carried out on our RTNN computer, which holds
32 PentiumPro processors, for a total CPU time of approximately one
month of the whole machine.
The errors in the estimation of the observables have been calculated
with the jackknife method.

For each of the two triplets $(\lambda_1,\lambda_2,\lambda)$,
we started by drawing the approximate
phase diagram in the plane $(\kappa_1,\kappa_2)$
for $T=0$, namely using symmetric lattices. This was done by simulating
an $8^3$ lattice and using hysteresis cycles to roughly identify
the transition lines. Afterwards, in
correspondence with each of these
approximate critical points, we performed on the same
lattice a simulation
with much larger statistics obtaining a more accurate estimate of the
transition as the maximum of the derivative along one of the axis
of the Binder cumulant defined below.
The resulting phase diagram for $\lambda=-0.016$
is shown in fig. \ref{PH_DIA}: we observe that the $(\kappa_1,\kappa_2)$
plane is divided in four
regions corresponding to the four possible cases of
$\langle\phi_1\rangle$ and $\langle\phi_2\rangle$
being zero or different from zero. For $\lambda=0$, the critical
lines would have been two straight lines each parallel to a
coordinate axis. The interaction among the fields curved
them in such a way that if one moves along, say, a vertical
line parallel to the $\kappa_2$ axis, starting from a point in
the disordered phase, one encounters first the $\phi_2$ transition;
if the value of $\kappa_2$ is now further increased, one eventually
crosses also the transition line for the $\phi_1$ field. This fact has
an intuitive explanation: after the field $\phi_2$ has undergone
the phase transition, it acquires a non-vanishing vev which, through
the quartic {\it negative} coupling with the $\phi_1$ field, acts on it
as an effective {\it negative} mass term. Increasing $\kappa_2$
makes $\langle \phi_2 \rangle$ increase, until when the induced
mass term for $\phi_1$ becomes large enough to cause a phase
transition for the latter field too. A similar behavior is observed
along lines parallel to the $\kappa_1$ axis.

Having got an idea of the phase diagram, we passed to the accurate
determination of some critical points.
Since they had to be
found with a high numerical precision, in order to clearly distinguish them
from those of the asymmetric lattices, we could not afford to explore
the entire $(\kappa_1,\kappa_2)$ plane, and thus we fixed once and for all
the value of $\kappa_1=0.33$, and searched on that vertical line
for the accurate critical values of $\kappa_2$,
$\kappa_2^c$, corresponding to the
transitions of $\phi_2$ and $\phi_1$.
For $\lambda=-0.016$ we simulated symmetric lattices
having $N_s=8,12,16,24,32$, while for  $\lambda=-0.020$ we had
$N_s=6,8,12,16$ and could not use larger lattices due to the large
correlation times. In any case, the results were very stable with $N_s$ and
larger values were not needed.
For each of the points and lattices that were simulated,
the combination of the Metropolis algorithm
with the Wolff single cluster method described above
was repeated $1,250,000$ times, taking measures every
$5$ iterations,
while in the case of the first order transitions sometimes
we made up to $5$ millions iterations.

The estimators used
and the method followed to take the thermodynamic limit, in the case
of second order phase transitions, were exactly the same as those
used in ref. \cite{bitu} and we refer to it the reader for the details.
Here we just give a short review.
To identify the critical points, we took the approximate values of
$\kappa_2^c$ given by the  $8^3$ lattice, and in
correspondence with them we measured
the Binder cumulants relative to that of the two fields which
was undergoing the phase transition. We recall that the Binder
cumulant of the field $\phi_i$ is defined as:
\be
U_{N_s,i}(\kappa_2)=\frac{3}{2}-
\frac{\langle M_i^4 \rangle}{2\langle M_i^2\rangle^2}~,\label{bin}
\ee
where
\be
M_i=\vert \frac{1}{V}\sum_{\sbbox{r}}\phi_{i,\sbbox{r}}\vert~.
\ee
Having measured the Binder cumulants, we extrapolated them in a narrow
$\kappa_2$ interval around the simulation point using the spectral
density method.
The values of $\kappa_2^c$ in the thermodynamic limit were obtained by
looking at the intersections $\kappa^*_2(N_{s1}, N_{s2})$ among all possible
pairs of curves $U_{N_s, i}(\kappa_2)$
and using the following scaling law
\cite{bin}:
\be
\kappa^*_2(N_{s}, b N_{s})- \kappa_2^c =
\frac{1-b^{-\omega}}{b^{1/\nu}-1} N_s^{-\omega -1/\nu}~,\label{binsca}
\ee
where $\omega$ is the exponent for the corrections to scaling. On the
symmetric lattices we used the exponents of the Ising model in three
dimensions, namely $\nu=0.63$, $\omega=0.8$, obtaining satisfactory fits.

\begin{figure}
\epsfig{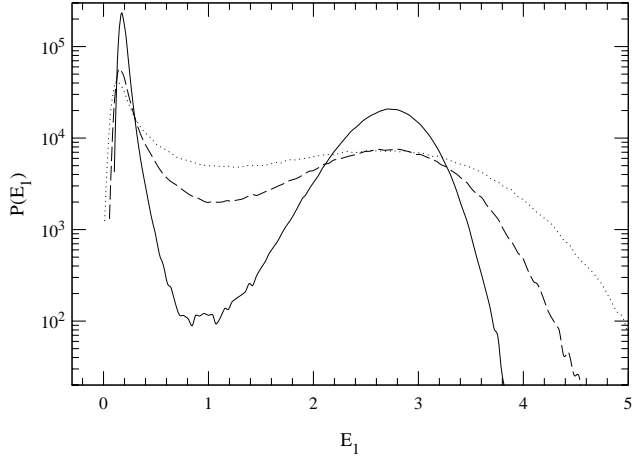}
\caption{Hystogram of $E_1$ for $\lambda=-0.020$ on the
$\phi_1$ transition
(the dotted, dashed and solid lines refer to the $6^3$, $8^3$ and
the $12^3$ lattices respectively).}
\label{histogramas}
\end{figure}

\begin{figure}
\epsfig{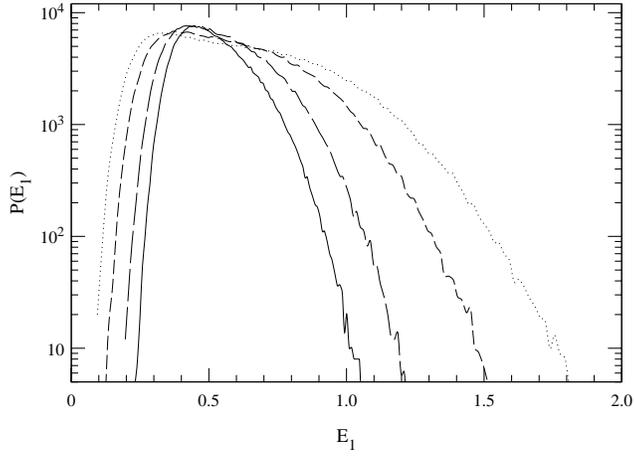}
\caption{Hystogram of $E_1$ for $\lambda=-0.016$ on the $\phi_1$
transition
($N_t=3$, dots for $N_s=12$, dashes $N_s=16$, daashes $N_s=24$, solid
$N_s=32$).}
\label{histog16}
\end{figure}

\begin{figure}
\epsfig{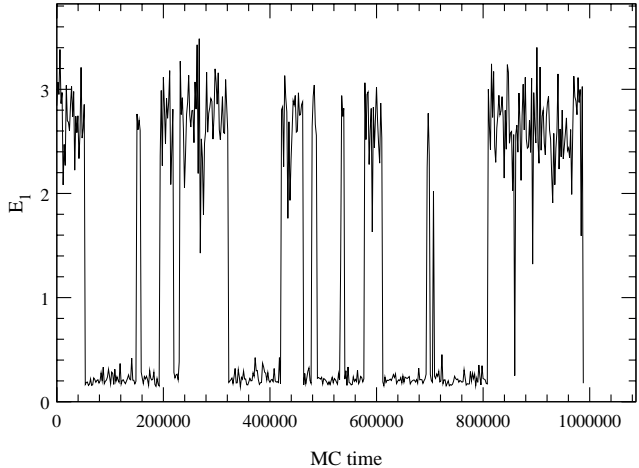}
\caption{Monte Carlo evolution of $E_1$ for $\lambda=-0.020$ on the
$\phi_1$ transition for the $12^3$ lattice.}
\label{evolucion}
\end{figure}

In the case $\lambda=-0.020$, the shape of the phase diagram
is very similar to that of $\lambda=-0.016$, except for one very significant
difference: while the transition of $\phi_2$ is still of
second order, that of $\phi_1$
turned out to be of {\it first} order (for $\lambda=-0.016$
both transitions were of second order).
To determine the precise location
of this transition, we did not use the same method as for the
second order ones. For first order phase transitions, general arguments of
FSS do not apply and, to be specific, the crossings of the Binder cumulants
do not coincide with the critical point.
In this case, we
searched for the value of $\kappa_2$, $\kappa_2(N_s)$, giving a maximum
of the specific heat:
\be
C_1=\frac{\partial E_1}{\partial \kappa_2}~,
\ee
where $E_1$ is the $\kappa_1$-energy of the field $\phi_1$:
\be
E_1=\frac{1}{Vd}\langle
\sum_{\sbbox{r} \in {\bf Z^3},\mu}
\phi_{1,\sbbox{r}}\phi_{1,\sbbox{r+\mu}}
\rangle,
\ee
which is known to give a good signal for first order phase
transitions. Using this estimator, the first order character
of this transition appears to be very clear:
fig. \ref{histogramas} shows how
the double peak structure, characteristic of first order transitions,
becomes stronger as $N_s$ increases. The figure
refers to symmetric lattices, but for all $N_t$ we have observed
a similar behavior. This figure can be compared with fig. \ref{histog16}
corresponding to the $\phi_1$ transition at $\lambda=-0.016$ and $N_t=3$
as an example (here it is apparent the second order character).
In fig. \ref{evolucion} we show a typical Monte Carlo evolution for
the first order case.

\begin{figure}
\epsfig{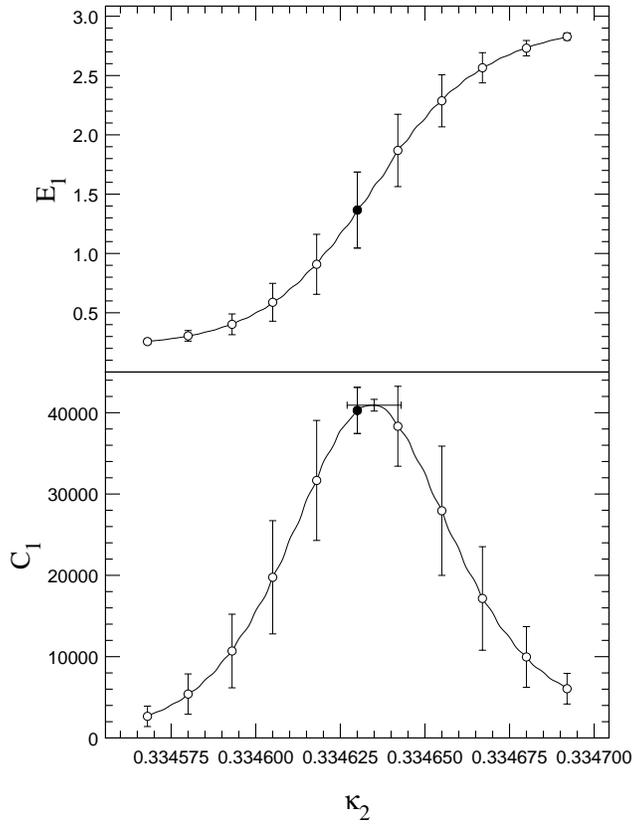}
\caption{$E_1$ and $C_1$ for $\lambda=-0.020$ on the
$\phi_1$ transition for the $12^3$ lattice by using the spectral density
method. The full mark denotes the simulation point, the empty ones are
extrapolations thereof.}
\label{ene_cal}
\end{figure}

Figure
\ref{ene_cal} gives an example of the $\kappa_1$-energy
and of the
corresponding specific heat, extrapolated from the simulation point
by means of the spectral density method.

The critical coupling in the
thermodynamical limit, i.e. $\kappa_2^{c}$, has been
obtained from the measurements of  $\kappa_2(N_s)$ by means of a fit
\be
\kappa_2(N_s)-\kappa_2^{c} \sim {N_s}^{-1/\nu}~,\label{firsca}
\ee
using for the critical exponent $\nu$ its value for first
order phase transitions in three dimensions,
$\nu=1/3$.
The fits turned out to be good in all cases, supporting the first
order character of the transition.

Having determined the phase diagrams at $T=0$, we passed to
the study of
the finite temperature case, by simulating asymmetric lattices
with $N_t < N_s$. The values
of $N_t$ that we considered have been $N_t=2,3,4$ for both values
of $\lambda$ and $N_t=5$ in one case. The values of $N_s$ used
were the same as for the $T=0$ case, except that for $\lambda=-0.020$
we have sometimes gone up to $N_s=24$.
The order of the transitions
turned out to be always the same as in the corresponding $T=0$
case; in particular, for all $N_t$'s we have found the first-order
transitions described above for the lower value of $\lambda$. The
techniques used to take the thermodynamic limit $N_s$ were the
same used in the $T=0$ case, except for two differences.
Since the
scaling parameter was $N_s$, while $N_t$ was fixed, according to the
hypothesis of dimensional reduction and universality, we  used
the exponents of the Ising model in {\it two} dimensions,
namely $\nu =1$ and $\omega=4/3$. Similarly, in the case of the
first order phase transition encountered for $\lambda=-0.020$,
in eq.(\ref{firsca}), we used the value of $\nu$ for two dimensions,
namely $\nu=1/2$. In both cases we got good fits.

\begin{figure}
\epsfig{figure= 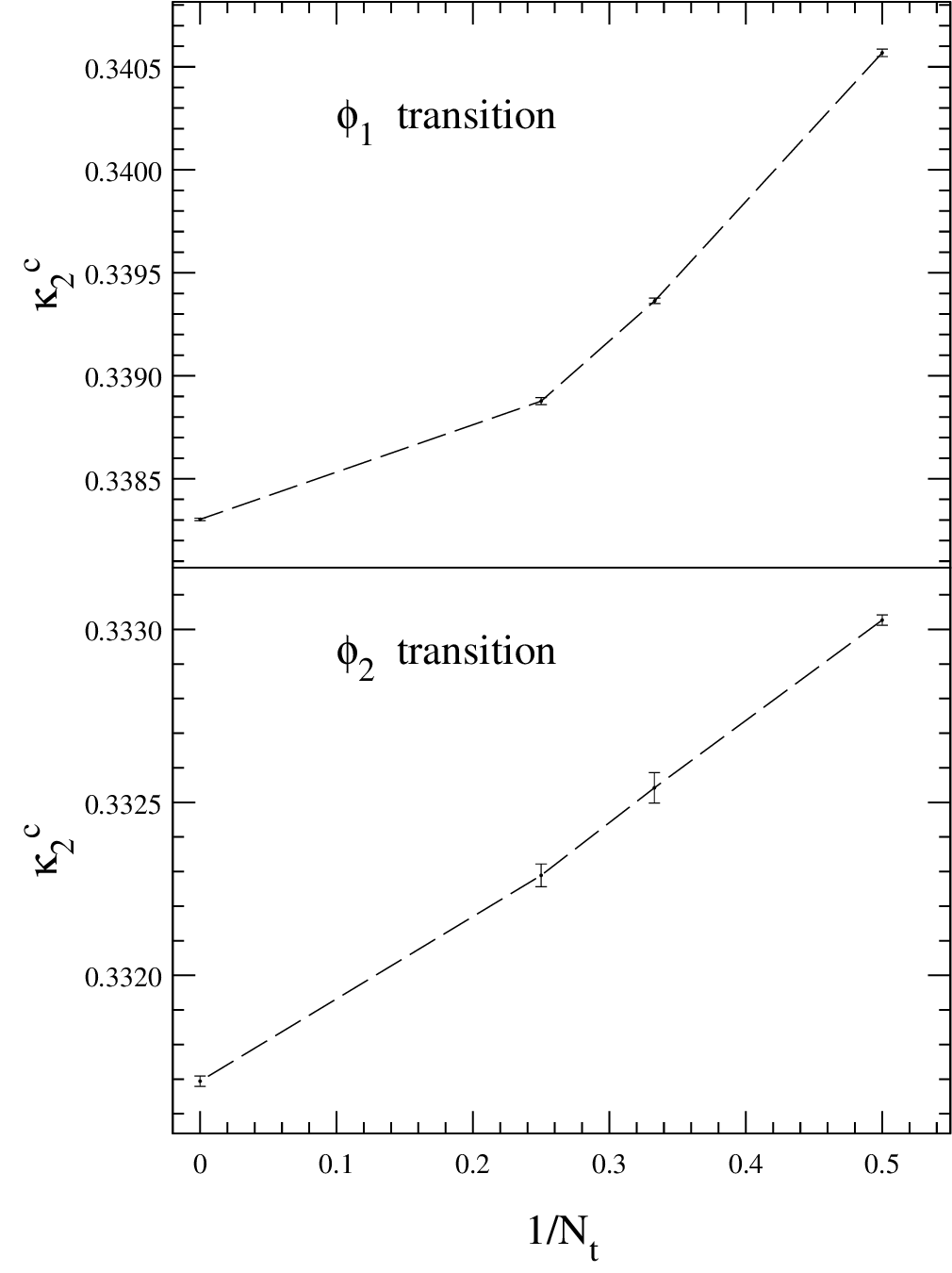,angle=0,width=240pt}
\caption{${\kappa_2}^{c}$ as a function of $1/N_t$ for $%
(\lambda_1=0.06,~\lambda_2=0.002, ~\lambda=-0.016,~ \kappa_1=0.33)$.
The points with $1/N_t$=0 refer to the symmetric lattices.}
\label{K2_NT_1}
\end{figure}
\begin{figure}
\epsfig{figure= 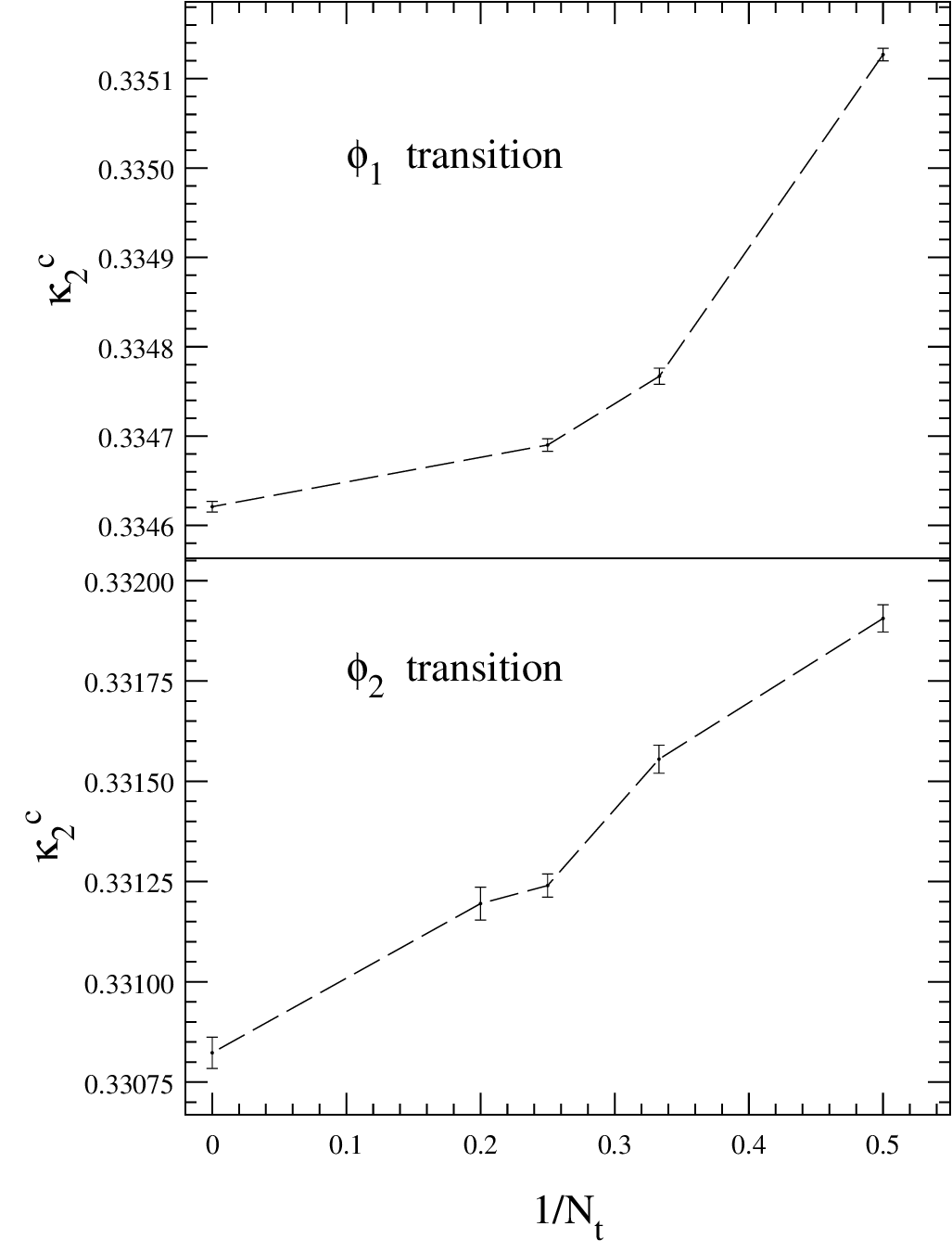,angle=0,width=240pt}
\caption{${\kappa_2}^{c}$ as a function of $1/N_t$ for $%
(\lambda_1=0.06,~ \lambda_2=0.002, ~\lambda=-0.02,~ \kappa_1=0.33)$.
The points with $1/N_t$=0 refer to the symmetric lattices.}
\label{K2_NT_2}
\end{figure}

As far as ISB is concerned,
the most important result of the simulations is apparent
from figs. \ref{K2_NT_1} and \ref{K2_NT_2}: for a constant $\kappa_1$,
they show that,
starting from the symmetric lattices (the points with
$1/N_t=0$),
$\kappa_2^{c}$ for the transition of the $\phi_2$ field
increases monotonically when $N_t$ is decreased for both
values of $\lambda$ considered;
this means that the critical points for $N_t$ finite shift deeper and
deeper in the
ordered region of the $T=0$ model, i.e. raising the temperature disorders
the system, as it happens in normal cases
(see for a comparison \cite{bitu})
and contrary to what is required for ISB to occur.
Though
we have considered a single value of $\kappa_1$ in our simulations
for finite $N_t$'s, we do not
expect a different behavior at other points. If for
some other value of $\kappa_1$
the  critical $N_t$-line passed below that for the $T=0$ theory,
they
would have to cross at some point, which would be a very strange fact.

With such a result, there is no need of looking at the continuum
limit to rule out ISB. Recalling the comments at the beginning of this
section,
it is obvious that the CCP's associated with
continuum theories having a symmetric vacuum at $T=0$ cannot
cross in any way the $\phi_2$ critical lines of the finite $N_t$ lattices,
for the latter lie in the {\it ordered} region of the phase
diagram for the symmetric lattices, and not in the {\it disordered} one.
Then this study indicates that Inverse Symmetry Breaking does not
work in this model.
However, it is not conclusive about Non Symmetry
Restoration because we do not know if the temperature for which the
symmetry is restored diverges or not in the continuum limit.

\section{Conclusions}

Inverse Symmetry Breaking  and Symmetry Non-Restoration
have recently attracted much interest,
mostly due do their potential implications for cosmological
scenarios. They are based on the possibility that in multiscalar
models some Debye masses can be negative at high temperatures, for
suitable choices of the couplings.

In this paper we have presented the results of a Monte Carlo study of ISB
in a simple model for two real scalars interacting via a
$\lambda\phi^2_1\phi^2_2$  coupling, in three euclidean dimensions. The
reasons for carrying such a study is that in three dimensions perturbations
theory cannot provide any information concerning these phenomena, as it is
discussed in Sec.3.  In order to have a fully non-perturbative approach to
these phenomena, we have thus simulated this scalar model on the lattice
and found no trace of ISB. It turned out that for all values of $\lambda$
that we considered the size of the disordered region of the phase diagram
increases, when the extension $N_t$ of the lattice in the euclidean-time
direction is decreased, as it occurs in normal systems, instead of
decreasing as required for ISB to occur. Heuristic arguments tell us that
this conclusion should be true for the whole parameter space. Another
interesting  result of our simulations is that, for large negative values
of $\lambda$, one of the transitions becomes of first order.

It would be desirable to know if other non-perturbative
analytical methods can reproduce our findings. Among the possible
approaches we mention here
the one based on the Average Effective Action \cite{wet},
or the CJT formalism \cite{corn}, that have already been used to
study the thermal behavior of two-scalar models in four dimensions
\cite{ame,roos,born2}.

We are currently simulating the model considered in this paper
in four dimensions, which is physically the most relevant case.
Our efforts are being focused on two issues, both very relevant
for Cosmology: one is the study of ISB
and SNR and the other is the existence of first order phase transitions
at finite temperature. We hope to report soon on this work.

\section*{Acknowledgments}

In our simulations we used the RTNN computer. This work has been
partially supported by CICYT under contract number
AEN96-1670. D.I. is a MEC (Spain) fellow
and C.L.U. is a DGA (Aragon, Spain) fellow.


\end{document}